\documentclass[aps,a4paper,10pt,twocolumn,nofootinbib,floatfix]{revtex4} %,twocolumn
\usepackage[T1]{fontenc}
\usepackage{pslatex}
\usepackage{amsmath}
\usepackage{amsfonts}
\usepackage{amssymb}
\usepackage{mathrsfs}
\usepackage[mathscr]{euscript}
\usepackage[dvips]{graphicx}
\usepackage{fancyhdr}
\usepackage{makeidx}
\usepackage{colordvi}

\def\overstrike#1#2{{\setbox0\hbox{$#2$}\hbox to \wd0{\hss
    $#1$\hss}\kern-\wd0\box0}}

\newcommand{\convol}{\star}
\newcommand{\grad}{\nabla}
\newcommand{\cross}{\times}

\begin{document}

\title{Optical pulse propagation with minimal approximations}
%\title{A first order wave equation for uni-directional pulse propagation}

\author{Paul Kinsler}
\email{Dr.Paul.Kinsler@physics.org}
\affiliation{
  Blackett Laboratory, Imperial College London,
  Prince Consort Road,
  London SW7 2AZ,
  United Kingdom.}

\begin{abstract}

Propagation equations for optical pulses
 are needed to assist in describing applications
 in ever more extreme situations -- 
 including those in metamaterials with linear and nonlinear magnetic responses.
Here I show how to derive a single first order propagation equation
 using a minimum of approximations
 and a straightforward ``factorization'' mathematical scheme.
The approach generates exact coupled bi-directional equations, 
 after which it is clear that the description 
 can be reduced to a single uni-directional first order wave equation
 by means of a simple ``slow evolution'' approximation, 
 where the optical pulse changes little over the distance 
 of one wavelength.
It also also allows a direct term-to-term comparison
 of an exact bi-directional theory with the approximate uni-directional theory.

\end{abstract}


%\PACS 42.25.Bs Wave propagation, transmission and absorption 
%\PACS 42.65.Re Ultrafast processes; optical pulse generation and 
%                 pulse compression 
%\PACS 31.15.-p Calculations and mathematical techniques in atomic 
%                 and molecular physics

% unused
%\PACS 78.20 Ci Optical constants (including refractive index, 
%                 complex dielectric constant, absorption, reflection 
%                 and transmission coefficients, emissivity) 

\lhead{\includegraphics[height=5mm,angle=0]{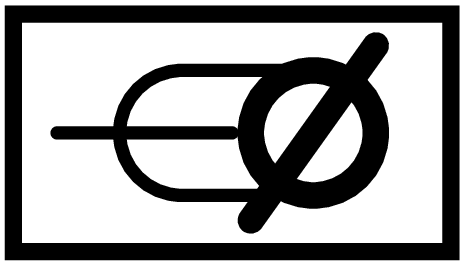}~~FCHHG}
\chead{Optical pulse propagation ...}
\rhead{
Dr.Paul.Kinsler@physics.org\\
http://www.kinsler.org/physics/
}
%\lfoot{(\yymmdddate\today:\currenttime) }
%\rfoot{{\large \emph{ Not for redistribution}}}

\date{\today}
\maketitle
\thispagestyle{fancy}

%
% ======================================================================
\section{Introduction}
\label{S-Introduction}

In recent years, 
 the propagation of optical pulses under ever more extreme conditions
 has been the subject of significant attention.
This situation has arisen primarily
 because of the multitude of applications \cite{Brabec-K-2000rmp}:
 e.g. where ultrashort pulses
 are relied upon to act as a kind of strobe-lamp
 to image ultrafast processes \cite{Corkum-1993prl,Schafer-YDK-1993prl},
 or where the electric field profile of a pulse
 \cite{Fuji-RGAUYTHK-2005njp,Radnor-CKN-2008pra}
 is engineered to excite specific atomic or molecular responses.
Other motivations are systems where
 strong nonlinearity is used to construct equally wide-band 
 but also temporally extended pulses -- 
 i.e. (white light) supercontinua 
 \cite{Alfano-S-1970prl,Alfano-S-1970prl-2,Dudley-GC-2006rmp} --
 or even come full circle and use the strong nonlinearity 
 to generate sub-structure that is again temporally confined,
 as in optical rogue waves \cite{Solli-RKJ-2007n}; 
 or even the temporally and spatially localized 
 filamentation processes \cite{Braun-KLDSM-1995ol,Chin-PBM-1999jjapl}.
Further, 
 developments in electromagnetic metamaterials
 \cite{Smith-PW-2004s, Litchinitser-S-2008lpl, Valentine-ZZUGBZ-2008n}
 lead to a requirement for including magnetic dispersion
 or even magnetic nonlinearity 
 \cite{Shadrivov-KWK-2008oe}.

It is clear, 
 therefore, 
 that progress toward shorter pulse durations
 as well as their increasing spectral bandwidths, 
 and higher pulse intensities --
 as well as exotic propagation media -- 
 are all factors
 either stretching existing pulse propagation models to their limits, 
 or breaking them.
In such regimes, 
 we need to be sure that our numerical models still work, 
 and have a clear idea of what has been neglected, 
 and what the side-effects of those approximations are.
Most existing pulse propagation models
 make sequential approximations that can have unforeseen side effects.
In contrast, 
 in this article, 
 I show how a straightforward and relatively simple derivation
 allows a side-by-side comparison of exact and approximate 
 propagation equations, 
 whilst still providing the numerical and analytical 
 convenience of a first-order wave equation.

The analysis of optical pulse propagation
 traditionally involves describing a pulse in terms
 of a complex field envelope, 
 while neglecting the underlying rapid oscillations at its carrier frequency. 
The resulting ``slowly varying envelope approximation'' (SVEA)  
 (see e.g. \cite{Shen-PNLO}), 
 which reduces second order differential equations to first order, 
 is valid when the envelope encompasses many cycles of the optical field
 and varies slowly.
Starting with the second order wave equation, 
 other auxiliary assumptions are required to get the final result 
 of a first-order wave equation:
 the introduction of a co-moving frame, 
 and the neglect of usually negligible second order spatial derivatives.
Although it is now easily possible to choose
 to solve Maxwell's equations numerically instead
 (see e.g. \cite{Flesch-PM-1996prl,Gilles-MV-1999pre,Brabec-K-1997prl,
                 Tarasishin-MZ-2001oc,Tyrrell-KN-2005jmo}), 
 the approach lacks the intuitive picture of a pulse ``envelope'', 
 and tends to be computationally demanding.

Many attempts have been made to generalize the SVEA style of derivation, 
 and perhaps the most notable of these was that of 
 Brabec and Krausz \cite{Brabec-K-1997prl}.
By slightly relaxing one assumption, 
 they derived corrections to the SVEA,
 which they included in their ``slowly evolving wave approximation'' (SEWA). 
This enabled the few-cycle regime to be modeled with improved accuracy, 
 and the SEWA has subsequently been applied in different situations, 
 including ultrashort IR laser pulses in fused silica 
 \cite{Ranka-G-1998ol,Tzortzakis-SFPMCB-2001prl},
 the filamentation of ultra-short laser pulses in air
 \cite{Akozbek-SBC-2001oc}, 
 and even in micro-structured optical fibres \cite{Gaeta-2002ol}. 
Later, 
 Porras \cite{Porras-1999pra} proposed a slightly different 
 ``slowly evolving envelope approximation'' (SEEA) that included
 corrections for the transverse behavior of the field; 
 and Kinsler and New  \cite{Kinsler-N-2003pra}
 took the process as far as it would go 
 with their ``generalized few-cycle envelope approximation'' (GFEA).
Although the wave equation generated by the GFEA was generally
 too complicated for practical use, 
 its derivation exposes one important point: 
 extending SVEA style derivations into wide-band situations 
 exposes the user to a number of poorly controlled side effects 
 \cite{Kinsler-2002-fcpp}.
Many other styles of dervivation also exist
 (see e.g. \cite{VanRoey-DL-1981josa,Scalora-C-1994oc,Geissler-TSSKB-1999prl}),
 but most use similar approximations,
 and apply them sequentially.

Here I will show that an alternative ``factorization'' style of derivation 
 we can achieve the simplicity of a first-order wave equation 
 for optical pulse propagation, 
 but avoid the unpleasant side-effects of the traditional approach.
Early but rather limited examples are by Shen \cite{Shen-PNLO}, 
 Blow and Wood \cite{Blow-W-1989jqe}, 
 and perhaps Husakou and Herrmann \cite{Husakou-H-2001prl}; 
 more recently (and more rigorously) we have 
 Ferrando et al. \cite{Ferrando-ZCBM-2005pre}
 and Genty et al. \cite{Genty-KKD-2007oe}.
The mathematical basis of the factorization shown in this article
 relies on Ferrando et al. \cite{Ferrando-ZCBM-2005pre}, 
 but here I make a point of generating wave equations
 incorporating most optical effects -- 
 both electric and magnetic dispersion, 
 diffraction, second and third order nonlinearity, 
 angle dependent refractive indices, and so on.
In particular,
 prior to any approximations being applied, 
 there is an (\emph{explicitly} bi-directional) stage 
 where two counter-propagating wave equations are coupled together.
This provides us with an important insight:
 that a simple ``slow evolution'' approximation is all that is needed
 to obtain a uni-directional first order wave equation, 
 irrespective of the origin of the coupling.

In this article I give a description of a modern
 approach to optical pulse propagation applicable to most situations
 that occur in nonlinear optics.
This is a regime where we want to model the most general situations possible, 
 while avoiding having to do a full numerical simulation
 of Maxwell's equations.
  The treatment here is intended to be straightforward enough for the student, 
   whilst also being comprehensive enough so that both novice and specialist
   can really understand the nature and limitations
   of this and other pulse propagation models.
Starting with a general form of the second order wave equation 
 in section \ref{S-secondorder}, 
 I follow with discussion the important role of the choice
 of propagation direction in section \ref{S-direction}, 
 which in nonlinear optics is usually in space and not in time.
In section \ref{S-factorisation}, 
 I introduce the method of factorization 
 that allows us to construct an explicitly bi-directional model, 
 and which is then reduced to the uni-directional limit 
 in section \ref{S-firstorder}, 
 where nonlinear pulse propagation is typically applied.
Section \ref{S-modifications} discusses typical modifications
 that can be applied to the equations 
 given in sections \ref{S-factorisation} and \ref{S-firstorder}
 in order to and simplify them appropriately 
 and compare them to existing models; 
 whilst section \ref{S-examples} gives specifc examples 
 for the common cases of propagation media 
 with second and third order nonlinearities.
The article is then summarized in section \ref{S-conclude}.

% ======================================================================
\section{Second order wave equation}
\label{S-secondorder}

Most optical pulse problems consider 
 a uniform and source free dielectric medium.
In such cases a good starting point 
 is the second order wave equation for the electric field, 
 which results from the substitution of the $\grad \cross \Vec{H} = \partial_t \Vec{D} + \Vec{J}$
 Maxwell's equation into the 
 $\grad \cross \Vec{E} =  -\partial_t \Vec{B}$ one
 (see e.g. \cite{Agrawal-NFO}), 
 although here I also allow for free currents $\Vec{J}$.
Magnetic effects can also be incorporated -- 
 easily so in the case of linear magnetic dispersion, 
 but also it is possible to retain a term for more general magnetic effects.
However, 
 cases where either the permittivity $\epsilon(\omega)$
 or the permeability $\mu(\omega)$ are negative are not excluded.

A sufficiently general model of the dielectric response
 in the time domain is
~
\begin{align}
  \Vec{D}(\Vec{r},t)
&=
  \epsilon(t)
 \convol
  \Vec{E}(\Vec{r},t)
\\
&=
  \epsilon_0
   \epsilon_L(\Vec{r},t)
  \convol
   \Vec{E}(\Vec{r},t)
 +
  \epsilon_0
  % \epsilon_L 
  %\convol
   \Vec{P}_\epsilon(\Vec{E},\Vec{r},t)
\label{eqn-DfromE}
,
\end{align}
 where the scalar $\epsilon_L$ contains the
 linear response of the material that is both \emph{isotropic}\footnote{The
  isotropy of $\epsilon_L$ (and later of $\mu_L$) 
  is both important and useful.}
 and lossless (or gain-less); 
 since here it is a time-response function, 
 it is convolved with the electric field $\Vec{E}$.
Note that the field vectors $\Vec{E}, \Vec{D}$, 
 and indeed the material parameter $\epsilon_L$
 are all functions of time $t$ and space $\Vec{r} = (x,y,z)$; 
 the polarization $\Vec{P}_\epsilon$ is a function
 of time $t$, 
 space $\Vec{r}$,
 and field $\Vec{E}$.
The following derivation also allows for magneto-electric polarizations, 
 i.e. those where $\Vec{P}_\epsilon$ also depends on $\Vec{H}$, 
 although I do not explicitly include such a dependence in the notation.
Similarly, 
 the magnetization response is 
~
\begin{align}
  \Vec{B}(\Vec{r},t)
&=
  \mu(t)
 \convol
  \Vec{H}(\Vec{r},t)
\\
&=
  \mu_0
   \mu_L(\Vec{r},t)
  \convol
   \Vec{H}(\Vec{r},t)
 +
  \mu_0
  % \epsilon_L 
  %\convol
   \Vec{M}_\mu(\Vec{H},\Vec{r},t)
\label{eqn-BfromH}
,
\end{align}
 where the scalar $\mu_L$ contains the
 linear response of the material that is both isotropic 
 and lossless (or gain-less).
Note that $\Vec{H}, \Vec{B}$ and $\mu_L$
 are all functions of time $t$ and space $\Vec{r} = (x,y,z)$; 
 the magnetization $\Vec{M}_\epsilon$ is a function
 of time $t$, 
 space $\Vec{r}$,
 and field $\Vec{H}$.
The following derivation also allows for magneto-electric magnetizations, 
 i.e. those where $\Vec{M}_\mu$ also depends on $\Vec{E}$, 
 although I do not explicitly include such a dependence in the notation.

Since here I have chosen to incorporate the ``simple'' linear responses
 of the propagation medium in $\epsilon_L$ and $\mu_L$, 
 the remaining parts $\Vec{P}_\epsilon$, $\Vec{M}_\mu$
 will usually be in part electric and magnetic field dependent, 
 and incorporate effects such as birefringence, 
 angle dependence,
 and nonlinearity; 
 it should also incorporate any loss \cite{Kinsler-2009pra}.
For example, 
 $\Vec{P}_\epsilon$ might contain a scalar nonlinearity
 such as third order Kerr nonlinearity
 with $P_{nl} \propto (\Vec{E} \cdot \Vec{E}) \Vec{E}$, 
 or a (vector) second order nonlinearity.
Note that it is not always necessary or desirable 
 to include all the simple linear responses
 in $\epsilon_L$ and $\mu_L$, 
 some may be left in $\Vec{P}_\epsilon$, $\Vec{M}_\mu$; 
 as will be discussed later.
Alternatively,  
 and in accordance with \cite{Kinsler-2009pra} we could choose
 to pick $\epsilon_L$ and $\mu_L$ such that $\epsilon_L \mu_L$ is real, 
 rather than each being real valued on its own.
However, 
 this would alter the handling of the $\Vec{J}$, 
 $\Vec{P}_\epsilon$, 
 and $\Vec{M}_\mu$ terms.

Defining 
 $\grad = (\partial_x, \partial_y, \partial_z)$
 and $\partial_a \equiv \partial / \partial a$, 
 $\epsilon_0 \mu_0 = 1/c^2$, 
 and current density $\Vec{J}$,
 we can write the exact second order wave equation as
~
\begin{align}
  c^2
  \grad
  \cross
  \grad
  \cross
  \Vec{E}(t)
&=
 -
  \partial_t^2
  \left[
    \mu_L(t)
    \convol
    \epsilon_L(t)
    \convol
    \Vec{E}(t)
  \right]
\nonumber
\\
& \quad~~
 -
  \partial_t^2
  \left[
    \mu_L(t)
    \convol
    \Vec{P}_\epsilon(t)
  \right]
 + 
  c^2 \partial_t \Vec{H}(t) \convol \left[ \cross \grad \mu_L(t) \right]
\nonumber
\\
& \qquad
 -
  \epsilon_0^{-1}
  \mu_L(t)
  \convol
  \partial_t 
  \Vec{J}
 -
  \partial_t
  \left[
   \frac{ \grad \cross \Vec{M}_\mu(t)}{\epsilon_0}
  \right]
.
\label{eqn-basic-nabla2E}
\end{align}
Here I have suppressed the space coordinates
 and electric field dependence for notational simplicity; 
 and will assume a homogeneous $\mu_L$ so that
 the $\partial_t \Vec{H} \cross \grad \mu_L$ term vanishes.
The (usual) next step is to replace 
 $\grad \cross \grad \cross \Vec{E}$ above 
 with the identity $\grad \grad \cdot \Vec{E} - \grad^2 \Vec{E}$, 
 where as usual $\grad^2 = \partial_x^2+\partial_y^2+\partial_z^2$.
Initially this might look over-complicated, 
 since $\grad \grad \cdot \Vec{E}$ adds in some extra terms
 (e.g. a $\partial_z^2 {E}_z$)
 which are then canceled by the same term from $\grad^2 \Vec{E}$.
However, 
 since the field divergence is an important Maxwell's equation, 
 splitting the double curl operation in this way turns out to be advantageous.

For the case of a free charge density $\rho$,
 and with the same separation of the material response
 as used above,
 Maxwell's equations tell us that 
~
\begin{align}
  \grad \cdot \Vec{D} 
=
 \rho
&= 
  \epsilon_0
  \grad 
  \cdot 
  \left[
     \epsilon_L
    \convol
     \Vec{E} 
 +
      \Vec{P}_\epsilon
  \right]
\\
&=
  \epsilon_0
     \epsilon_L
    \convol
  \grad 
  \cdot 
  \Vec{E} 
 +
  \epsilon_0
  \left[
    \grad 
     \epsilon_L
  \right]
    \cdot
    \convol
  \Vec{E} 
 +
   \epsilon_0
  \grad 
  \cdot 
      \Vec{P}_\epsilon
\\
  \Longrightarrow
  \epsilon_L \convol
    \grad
   \cdot
    \Vec{E}
&= 
 -
      \grad
   \cdot
     \Vec{P}_\epsilon
 -
  \rho
\label{eqn-factor-divD}
\end{align}
so for an isotropic and homogeneous $\epsilon_L$, 
 we can use $\grad \epsilon_L = 0$; 
 note that isotropy also implies field-independence.
The frequency domain changes convolutions into products, 
 so that we have
~
\begin{align}
  \epsilon_0
  \epsilon_L (\omega)
    \grad
   \cdot
    \Vec{E} (\omega)
&= 
  \rho (\omega)
 -
      \grad
   \cdot
     \Vec{P}_\epsilon (\omega)
\\
    \grad
   \cdot
    \Vec{E} (\omega)
&=
   \frac{\rho (\omega)}
        {\epsilon_0 \epsilon_L (\omega)}
 -
   \frac{\grad
         \cdot
         \Vec{P}_\epsilon (\omega)
        }
        {\epsilon_L (\omega)}
.
\label{eqn-factor-divD-iso}
\end{align}
Note that the left-hand side (LHS)
 of this equation (i.e. $\grad \cdot \Vec{E}$)
 seems to be potentially large, 
 since it consists of field derivatives.
However, 
 the divergence condition reveals that with no free charge
 it is simply $\grad \cdot \Vec{P}_\epsilon/\epsilon_L$, 
 which merely is of the order of the nonlinearity or anisotropy of $\epsilon$;  both of which are small in typical systems.
Since $\grad [\grad \cdot \Vec{E}]$ is typically 
 much smaller than $\grad^2 \Vec{E}$, 
 it can reasonably be considered as a correction
 to a propagation dominated by $\grad^2 \Vec{E}$.

As a result, 
 we find that the replacement of $\grad \cross \grad \cross \Vec{E}$
 by $-\grad^2 \Vec{E} + \grad \grad \cdot \Vec{E}$
 not only achieves this valuable minimization, 
 but it also reduces the remaining spatial derivatives 
 to the simple $\grad^2 \Vec{E}$.
The side effect is that we now need
 to compute $\grad \grad \cdot \Vec{P}_\epsilon$,
 which may well be a complicated function of $\Vec{E}$; 
 it also gives rise to phenomena such as nonlinear diffraction term
 (see e.g. \cite{Boardman-MPS-2000oqe}).

The second order wave equation
 is best written in the frequency domain, 
 because of the need to divide the divergence term
 by the frequency dependent $\epsilon_L$; 
 and so is
~
\begin{align}
 -
  c^2
  \grad^2
  \Vec{E}(\omega)
&=
  \omega^2
  \partial_t^2
  \epsilon_L(\omega)
  \mu_L(\omega)
  \Vec{E}(\omega)
 +
  \omega^2
  \mu_L(\omega)
  \Vec{P}_\epsilon(\omega)
\nonumber
\\
& \qquad
 +
  \imath 
  \frac{\omega}{\epsilon_0}
  \mu_L(\omega)
  \Vec{J}(\omega)
 +
  \imath 
  \frac{\omega}{\epsilon_0}
  \grad \cross \Vec{M}_\mu
\nonumber
\\
& \qquad \quad
 +
  \frac{c^2}
       {\epsilon_L(\omega)}
  \grad
  \left[
    \grad \cdot \Vec{P}_\epsilon(\omega)
   -
    \frac{\rho(\omega)} 
         {\epsilon_0}
  \right]
.
\label{eqn-basic-nabla2E-Helmholtz}
\end{align}
For plane polarized pulses, 
 a scalar version allowing for just one of the linear polarization
 components is sufficient. 
However for materials that couple
 the horizontal and perpendicular polarizations together,
 such as the $\chi^{(2)}$ interaction relied on by
 optical parametric amplifiers (OPA) or oscillators
 (see e.g. \cite{Boyd-NLO}),
 we could write one equation for each polarization, 
 and then find that they were coupled together
 by the nonlinearity.

The wave equation in eqn. \eqref{eqn-basic-nabla2E-Helmholtz} contains
 both current $\Vec{J}$ and charge density $\rho$ terms, 
 which are usually interdependent.
These terms are not often important in pulse propagation, 
 so I do not discuss their modeling; 
 appropriate treatments can be seen in the literature
 on optical filamentation (see e.g. \cite{Berge-S-2009dcdsa}).

% ======================================================================
\section{Propagation direction}
\label{S-direction}

In this article I will not be considering 
 strong reflections from material modulations or interfaces.
Nevertheless, 
 considering simple reflections is an excellent way
 of clarifying some important issues
 that arise when we choose whether to propagate pulses 
 forward in time, 
 or forward in space.

\begin{figure}
 \includegraphics[width=0.80\columnwidth,angle=0]{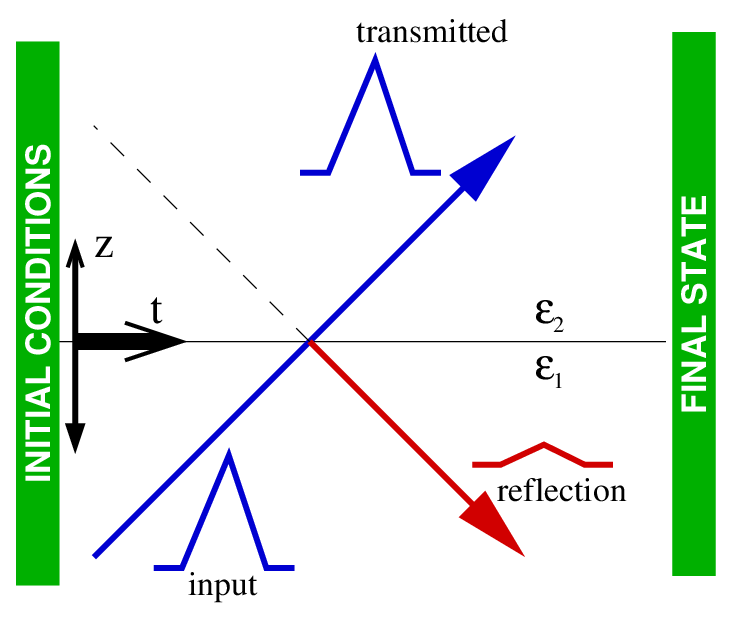}
\caption{
An ordinary reflection at an interface 
 between media with permittivities $\epsilon_1$ and $\epsilon_2$, 
 in a $t$-propagated picture.
An incoming pulse propagates forward (in $t$) 
 and evolves forward (in $z$) until it reaches an interface, 
 whereupon it splits into 
 a transmitted pulse and a normal reflected pulse; 
 the reflected pulse then evolves backward in space
 as both transmitted and reflected pulses continue
 to propagate forward in time.
}
\label{F-diagram-reflect-t}
\end{figure}

Temporal propagation is the usual choice in finite difference time domain
 (FDTD) modeling of Maxwell's equations \cite{Yee-1966tap,Joseph-T-1997itap}, 
 where fields $\Vec{E}(x,y,z), \Vec{H}(x,y,z)$
 are stepped forward in time $t$;
 exitations of the field (i.e. optical pulses)
 then evolves backward or forwards in the space coordinates $(x,y,z)$.
We therefore set up initial conditions covering each point in space
 at a chosen initial time $t_i$; 
 likewise we read out our final state for each point in space
 at a chosen final time $t_f$, 
 as shown on fig. \ref{F-diagram-reflect-t}.
This choice requires a time-response treatment of dispersion, 
 perhaps involving convolutions, 
 however,
 as also shown by fig. \ref{F-diagram-reflect-t}, 
 it provides natural reflections.

\begin{figure}
 \includegraphics[width=0.80\columnwidth,angle=0]{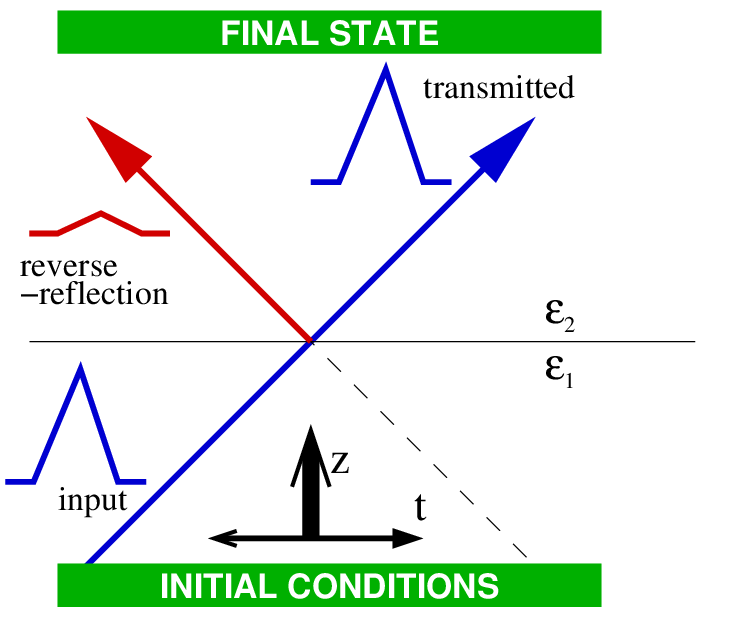}
\caption{
A reflection at an interface 
 between media with permittivities $\epsilon_1$ and $\epsilon_2$, 
 in a $z$-propagated picture.
An incoming pulse propagates forward (in $z$) 
 and evolves forward (in $t$) until it reaches an interface, 
 whereupon it splits into 
 a transmitted pulse and its reverse reflection;
 the reverse reflected pulse then evolves backward in time
 as both transmitted and reflected pulses continue
 to propagate forward in space.
A reverse reflection is the means by which a spatially propagated 
 system represents a pulse propagating backwards in $z$,
 so that it cancels the ordinary reflection
 missing from the initial conditions.
}
\label{F-diagram-reflect-z}
\end{figure}

Spatial propagation is the usual choice in nonlinear optics
 and optical pulse propagation, 
 where fields $\Vec{E}(t,x,y), \Vec{H}(t,x,y)$
 are stepped forward in a chosen spatial direction ($z$);
 exitations of the field (i.e. optical pulses)
 then evolves backward or forwards in time and space coordinates $(t,x,y)$.
We therefore set up initial conditions covering each point in time
 at a chosen point in space $z_i$;
 likewise we read out our final state for each point in time
 at a chosen point in space $z_f$, 
 as shown on fig. \ref{F-diagram-reflect-z}.
Comparison of figs. \ref{F-diagram-reflect-t} and \ref{F-diagram-reflect-z}
 also show that to be correctly modeled,
 an ordinary reflection from the interface back to our initial point 
 must be included in our initial conditions.
Unfortunately, 
 we will usually not know the properties of this reflection in advance, 
 so we will not include it in the initial conditions.
As a result,
 our solution of Maxwell's equations at the interface 
 creates the mirror image pulse that is needed 
 to exactly cancel out the ordinary reflection.
Next, 
 since we have chosen to propagate solely toward larger $z$, 
 this mirror image ```reverse reflection'' pulse
 now evolves forward in space $z$ but backwards in time $t$, 
 as shown on fig. \ref{F-diagram-reflect-z}.

We see, 
 therefore, 
 that if we want to take advantage of the benefits of spatial propagation, 
 notably the eficient handling of dispersion, 
 we will also not want to be modeling systems containing 
 significant reflections.
Indeed, 
 this issue motivated the time-propagated model
 of Scalora et al. 
 \cite{Dowling-SBB-1994jap,Scalora-DBB-1994prl,Scalora-C-1994oc}, 
 which are based on the second order wave equation; 
 however that approach suffers some of the same drawbacks
 as other tradition pulse propagation techniques.
To handle a temporally propagated model based on a second order wave equation, 
 it is best to use that for the displacement field $\Vec{D}$
 rather than for $\Vec{E}$; 
 since time derivatives of $\Vec{D}$ appear directly
 in Maxwell's equations, 
 whereas those for $\Vec{E}$ are complicated by the material response.

%
% ----------------------------------------------------------------------
\subsection{Spatial propagation}
\label{S-direction-z}

The first step to achieving a first order wave equation containing
 the necessary physics but without unnecessarily complex approximations
 is to reorganize the wave eqn. \eqref{eqn-basic-nabla2E}
 to emphasize contributions that by themselves 
 can freely propagate forward and backward without interacting.
To do this I choose
 a specific propagation direction (e.g. along the $z$-axis), 
 and then denote the orthogonal components (i.e. along $x$ and $y$) 
 as transverse behaviour; 
 many situations are also cylindrically symmetric, 
 allowing simplification of the two transverse dimensions $x, y$
 into a single radial coordinate $r$.
I therefore rearrange eqn. \eqref{eqn-basic-nabla2E-Helmholtz} 
 into
~
\begin{align}
  \left[
    \partial_z^2
  +
    \beta^2(\omega)
  \right]
    \Vec{E}(\omega)
&=
 -
    \grad_\perp^2
    \Vec{E}(\omega)
 -
  k_0^2
  \mu_L
  \Vec{P}_\epsilon(\omega)
\nonumber
\\
& \quad
 -
  \imath
  c k_0 \mu_0 \mu_L 
  \Vec{J}(\omega)
 -
  \imath
  c k_0 \mu_0
  \grad \cross \Vec{M}_\mu
\nonumber
\\
& \quad
 -
  \frac{1}
       {\epsilon_L(\omega)}
  \grad
  \left[
    \grad \cdot \Vec{P}_\epsilon(\omega)
   -
    \frac{\rho(\omega)}
         {\epsilon_0}
  \right]
,
\label{eqn-bi-d2zE}
\end{align}
where $k_0^2 = \omega^2 / c^2$
 and $\beta^2(\omega) = k_0^2 n^2
  = \omega^2 \epsilon_0 \mu_0 \epsilon_L(\omega) \mu_L(\omega)$;
 $k_0^2 = \omega^2 / c^2$.
Here all the simple linear response 
 (e.g. the isotropic refractive index and dispersion)
 has been moved to the LHS
 as a (possibly) frequency dependent propagation wave vector;
 the residual responses (i.e. $\Vec{P}_\epsilon$ and $\Vec{M}_\mu$)
 contain
 any non-$\omega$ dependence,
 angle dependent terms, 
 nonlinearity
 or spatial variation.
Note that defining $\beta(\omega)$ is a matter of \emph{choice}, 
 in some cases we may find it convenient to define it
 to be frequency independent; 
 in others we might (e.g.) even decide to retain some angle dependence, 
 perhaps even to the point of generating
 a spherical ``in-out'' bi-directional model, 
 rather than a linear forward-backward one.

%
% ======================================================================
\section{Factorization}
\label{S-factorisation}

I now factorize the wave equation, 
 a process which, 
 while used in optics for some time \cite{Blow-W-1989jqe}
 has only recently been used to its full potential
 \cite{Ferrando-ZCBM-2005pre,Genty-KKD-2007oe,Kinsler-2007josab}.
Factorization neatly avoids almost all of the approximations
 necessary in the standard approach and its extensions
 \cite{Boyd-NLO,Brabec-K-1997prl,Geissler-TSSKB-1999prl,Kinsler-N-2003pra}
 (etc) --
 which are in fact much more complicated than they first appear, 
 as has been shown by detailed analysis 
 \cite{Kinsler-N-2003pra,Kinsler-2002-fcpp}.
A major advantage of factorization is that we can 
 directly compare the exact bi-directional
 and approximate uni-directional theories term for term, 
 whereas in other approaches the backward parts simply vanish
 and are not directly available for comparison.
Perhaps the clearest recent description of the approximations
 made in a standard 
 (non-factorization) derivation
 of a uni-directional wave equation
 is by Berge and Skupin \cite{Berge-S-2009dcdsa}.
That work discussed the filamentation
 resulting from nonlinear self-focusing effects, 
 so that they incorporated the role
 of the longitudinal field components
 and included a model for a plasma 
 connecting the $\Vec{J}$ and $\rho$ contributions --
 here I retain these terms, 
 but the reader is referred to Berge and Skupin \cite{Berge-S-2009dcdsa}
 for a specific model.

Factorization takes its name from the fact that 
 the LHS of eqn. \eqref{eqn-bi-d2zE} % {eqn-bi-d2zEperp}
 is a simple sum of squares which might be factorized, 
 indeed this is what was done in 1989 in a somewhat ad hoc fashion 
 by Blow and Wood \cite{Blow-W-1989jqe}.
Since the factors are just $(\partial_z \mp \imath \beta)$, 
 each by itself looks like a forward (or backward) directed wave equation.
A rigorous factorization procedure
 \cite{Ferrando-ZCBM-2005pre,Kinsler-2007-envel},
 of which some basics are given in appendix \ref{S-factorize}, 
 allows us to define a pair of counter-propagating Greens functions, 
 and so divide the second order wave equation
 into a bi-directional pair of coupled 
 first order wave equations.
That these factorized equations are equivalent to the original
 second order wave equation is proven by taking their sum and differences, 
 then substituting one into another with the assistance
 of a derivative with respect to $z$, 
 as explained in Ref. \cite{Ferrando-ZCBM-2005pre}\footnote{See 
  section IV.B of this reference.}.
Further, 
 even in the approximate uni-directional limit, 
 the factorised wave equations have been shown
 by Genty et al. \cite{Genty-KKD-2007oe} to give
 a stunning level of agreement
 with pseudospectral spatial domain (PSSD) \cite{Tyrrell-KN-2005jmo}
 Maxwell equations simulations.

Before proceeding, 
 it is worth reiterating an important point --
 the choice of $\epsilon_L(\omega)$ and $\mu_L(\omega)$, 
 and therefore of $\beta(\omega)$
 in eqn. \eqref{eqn-bi-d2zE}, % (or \eqref{eqn-bi-d2zEperp}) 
 defines the specific Greens functions used; 
 it therefore also defines the underlying basis 
 on which we will then propagate the electric field $\Vec{E}$.

As an aside, 
 the interested reader may wish to examine the mathematical
 ``wave-splitting'' work of Weston and others
 (see e.g. \cite{Weston-1993jmp}), 
 although it does not consider residual terms, 
 and (at least initially) was primarily concerned
 only with reflections and scattering.
This was based on that from the earlier work of 
 Beezley and Krueger \cite{Beezley-K-1985jmp} 
 who applied wave-splitting concepts to optics.
Other similar work is the one-way wave equation
 of Leviandier \cite{Leviandier-2009jpa}, 
 and other directional schemes have been suggested
 by Kinsler et al. \cite{Kinsler-RN-2005pra}
 and Kolesik et al. \cite{Kolesik-MM-2002prl}.
It is also interesting to compare and contrast 
 the factorization scheme used here with
 beam propagation methods 
 (BPM, e.g., Refs. \cite{VanRoey-DL-1981josa,Scarmozzino-GPH-2000stqe}).
For example, 
 the treatment of Van Roey et al. \cite{VanRoey-DL-1981josa} 
 also begins using Greens' functions 
 which define a chosen reference propagation.
Thus, 
 whilst those BPM methods might (in principle)
 be developed in a way which matches 
 the benefits of the factorization scheme I present here, 
 to my knowledge no such implementation has been published.

%
% ----------------------------------------------------------------------
\subsection{Bi-directional wave equations}

A pair of bi-directional wave equations suggests 
 similarly bi-directional fields, 
 so I split the electric field 
 into forward ($\Vec{E}^+$) and backward ($\Vec{E}^-$) directed parts, 
 with $\Vec{E} = \Vec{E}^+ + \Vec{E}^-$.
In the following equations 
 I have reinstated the $\Vec{E}$ argument of $\Vec{P}_\epsilon$
 (and $\Vec{H}$ of $\Vec{M}_\mu$)
 to emphasise that they depend 
 on the total field;
 an important point since we see that $\Vec{P}_\epsilon$, 
 $\Vec{M}_\mu$, 
 diffraction, 
 and other terms 
 drive both forward and backward equations equally.

Using the procedure summarized in Appendix \ref{S-factorize}, 
 the second order wave equation in eqn. \eqref{eqn-bi-d2zE}
 can be converted into a pair of 
 coupled bi-directional first order wave equations
 for the directed fields $\Vec{E}^\pm$.
They are
~
\begin{align}
    \partial_z
  \Vec{E}^\pm(\omega)
&=
 \pm
    \imath
    \beta(\omega)
  \Vec{E}^\pm(\omega)
%\nonumber
%\\
%& 
 ~~
 \pm
  \frac{\imath \grad_\perp^2}{2 \beta(\omega)}
  \left[
    \Vec{E}^+(\omega) + \Vec{E}^-(\omega)
  \right]
\nonumber
\\
&\quad~~
 \pm
  \frac{\imath k_0^2(\omega) \mu_L}
       {2 \beta(\omega)}
  \Vec{P}_\epsilon
  (
    \Vec{E}^+ + \Vec{E}^-, {E}_z, \omega
  )
\nonumber
\\
& \qquad
 \mp
  \frac{c k_0(\omega) \mu_0 \mu_L}
       {2 \beta(\omega)}
  \Vec{J}(\omega)
 \mp
  \frac{c k_0(\omega) \mu_0}
       {2 \beta(\omega)}
  \grad \cross \Vec{M}_\mu
\nonumber
\\
& \qquad~~
 \pm
  \frac{\imath}
       {2 \beta(\omega) \epsilon_L(\omega)}
  \grad
  \left[
    \grad \cdot \Vec{P}_\epsilon(\omega)
   -
    \frac{\rho(\omega)}
         {\epsilon_0}
  \right]
.
\label{eqn-bi-dzE}
\end{align}
Since $k_0 = \omega/c$, 
 such factors convert to a (scaled) time derivative
 when these frequency domain equations
 are transformed into the time domain.

%
% ----------------------------------------------------------------------
\subsection{Propagation, evolution, and directed fields}
\label{S-factorisation-propevodir}

Note that since our solutions of the wave equations 
 enforce \emph{propagation} toward larger $z$, 
 the fields ${E}^\pm(t)$ are \emph{directed} forwards and backward in time; 
 these fields then \emph{evolve} 
 forwards and/or backward in time as $z$ increases.
Note that I use this terminology 
 (propagated, directed, evolved)
 throughout this article
 to mean these three specific and distinct concepts.

%%
%% ----------------------------------------------------------------------
%\subsection{Contributions to the evolution}
%\label{S-factorisation-contributions}

When examining the wave equation eqn. \eqref{eqn-bi-dzE}
 which evolves the directed fields ${E}^\pm$
 as they propagate forward in $z$, 
 we see that the right-hand side (RHS) has two types of terms:
 which I label the ``underlying'' 
 and ``residual'' parts \cite{Kinsler-2009pra}.

\emph{Underlying evolution} 
 is that given by $\pm \imath \beta(\omega) {E}^\pm$ term, 
 and is determined by our chosen $\epsilon_L(\omega)$ and $\mu_L(\omega)$.
By itself, 
 it would describe a plane-wave like evolution where the 
 field oscillations would move forward ($+$) or backward ($-$)
 in time across ${E}^\pm(t)$.
This is analogous to the choice of reference when 
 constructing directional fields \cite{Kinsler-RN-2005pra}, 
 or the refractive index term $n_0^2$ 
 used in the BPM \cite{VanRoey-DL-1981josa}.

\emph{Residual evolution}
 accounts for the discrepancy between the true evolution
 and the underlying evolution, 
 and is every part of the material response \emph{not} 
 included in $\epsilon_L(\omega)$ or $\mu_L(\omega)$; 
 i.e. it is all the remaining terms on the RHS
 of eqn. \eqref{eqn-bi-dzE}.
These typically include any non-linear polarization, 
 angle dependent linear terms, 
 and the transverse effects; 
 they are analogous to the correction terms 
 used in directional fields models, 
 or the refractive index perturbation $\Delta n^2$ 
 used in BPM \cite{VanRoey-DL-1981josa}.
In the language used by Ferrando et al. \cite{Ferrando-ZCBM-2005pre}, 
 these residuals are ``source'' terms.
Although we might hope they will be a weak perturbation, 
 so that we could make the (desirable) uni-directional approximation
 discussed later, 
 the factorization procedure is valid for \emph{any} strength.

%
% ----------------------------------------------------------------------
\subsection{Underlying evolution: 
              choice of $\beta$ and the resulting ${E}^\pm$}
\label{S-factorisation-beta}

I now examine how the choice of $\beta$ affects the 
 relative sizes of the forward and backward directed ${E}^+$ and ${E}^-$.
To do this I consider the simple example of a medium for which the field 
 is known to propagate with wave vector $k$; 
 but for demonstration purposes we choose an underlying evolution 
 determined by a wave vector $\beta$
 that is different from $k$.
For example,
 for a linear isotropic medium we could exactly define
 $k^2 = \beta^2 + \Delta^2$; 
 but in general we would just have some residual (source) term $\mathscr{Q}$.
This means that our definitions of forward and backward \emph{directed} fields
 do not exactly correspond to what the wave equation
 will actually \emph{evolve} forward and backward
 as we \emph{propagate} toward larger $z$.

The second order wave equation is
 $(\partial_z^2+\beta^2) {E} = - \mathscr{Q}$, 
 which in the linear case has $\mathscr{Q}=\delta^2 {E}$, 
 so that $(\partial_z^2+k^2) {E} = 0$.
The factorization
 in terms of $\beta$ is then
~
\begin{align}
  \partial_z
  {E}^\pm
&=
 \pm
  \imath \beta {E}^\pm
 \pm
  \frac{\imath \mathscr{Q}}{2 \beta}
.
\label{eqn-choosebeta-factor}
\end{align}
Now if we select the case where our field ${E}$ only evolves forward, 
 we know that ${E} = {E}_0 \exp [\imath k z]$.
Consequently ${E}^\pm$ must have matching oscillations: 
 i.e. ${E}^\pm = {E}_0^\pm \exp [\imath k z]$,
 even though ${E}^-$ is directed backward.
Substituting these 
 into eqn. \eqref{eqn-choosebeta-factor} gives
~
\begin{align}
  {E}_0^-
&=
  \frac{\beta - k}
       {\beta + k}
  {E}_0^+
,
\label{eqn-choosebeta-EmEp}
\end{align}
 which specifies how much ${E}^-$ we need to combine with ${E}^+$
 so that our pulse evolves forward; 
 since the ${E}^-$ will be dragged forward by its coupling to ${E}^+$.
This interdependent ${E}^\pm$ behaviour is generic -- 
 no matter what the origin of the discrepancy between $\beta$
 and the true evolution of the field
 (i.e. the residual or source terms such as 
  mismatched dispersion, nonlinearity, diffraction, etc):  
 some non-zero backward directed field ${E}^-$ must exist
 but still evolve forwards with ${E}^+$.
Analogous behaviour can be seen in the directional fields
 approach of Kinsler et al. \cite{Kinsler-RN-2005pra}.

Usually we hope that this residual ${E}^-$ contribution
 is small enough so that it can be neglected.
If we assume ${E}^- \simeq 0$, 
 then we find that $k \simeq \beta + \Delta^2/2\beta$, 
 which is just the expansion of $k = (\beta^2 + \Delta^2)^{1/2}$
 to first order in $\Delta^2/\beta^2$.
Following this, 
 we find that eqn. \eqref{eqn-choosebeta-EmEp} then says that 
 ${E}_0^- \simeq (\Delta^2/4\beta^2) {E}_0^+$, 
 which has come full circle and provided us with the scale
 on which ${E}^-$ can be considered negligible.
Outside the restricted (linear) case where we know $\Delta^2$, 
 the true wave vector $k$ might be difficult to determine, 
 and in nonlinear propagation may even change as the pulse propagates.

There is a further important point to notice:
 if we choose $\beta=\beta(\omega)$ with a frequency dependence, 
 then we see that the source-like terms
 (e.g. diffraction, polarization, etc; 
  or $\Delta^2$ in eqn.  \eqref{eqn-choosebeta-factor}) 
 inherit that dispersion.
This means that even if we started with polarization model 
 with instantaneous nonlinearity, 
 our factorized equations no longer have instantaneous nonlinear terms, 
 as they have become ``anti-dispersed'' by the factor of $\beta(\omega)^{-1}$; 
 as indeed have the other residual terms.
This matches exactly what happens in the directional fields
 approach of Kinsler et al. \cite{Kinsler-RN-2005pra}, 
 where choosing a dispersive reference has an equivalent effect
 on the correction terms.

%
% ======================================================================
\section{Uni-directional wave equations}
\label{S-firstorder}

Making only a single
 well defined type of approximation I can now reduce the
 exact coupled bi-directional evolution of $\Vec{E}_\perp$ down to a 
 single uni-directional 
 first order wave equation.
I do not require a moving frame, 
 a smooth envelope, 
 or to assume inconvenient second order derivatives 
 are somehow negligible:
 all these are frequently required in standard treatments, 
 and even extensions use them
 \cite{Boyd-NLO,Brabec-K-1997prl,Porras-1999pra,Kinsler-N-2003pra,
        Geissler-TSSKB-1999prl,Scalora-C-1994oc}.
The approximation is that the residual terms are weak
 compared to the (underlying) $\pm \imath \beta \Vec{E}$ term --
 e.g. weak nonlinearity, 
 angle dependence, 
 and diffraction.
This enables me to assert that if I start with $\Vec{E}^-=0$, 
 then $\Vec{E}^-$ will remain negligible --
 see my estimate in subsection \ref{S-factorisation-beta}.
In this context, 
 ``weak'' means that no significant change in the backward field 
 is generated in a distance shorter than one wave period
 (``slow evolution''); 
 and that small effects do not build up gradually over propagation distances 
 of many wavelengths (``no accumulation'').

\emph{Slow evolution} 
 is where the size of the residual terms 
 is much smaller than that of the underlying linear evolution --
 i.e. smaller than $\beta \Vec{E}$.
This allows us to write down straightforward inequalities
 which need to be satisfied.
It is important to note the close relationship between these and 
 a good choice of $\beta$, 
 as discussed in subsection \ref{S-factorisation-beta}.
If $\beta$ is not a good enough match, 
 there always be significant contributions from both
 forward and backward directed fields; 
 and even if nothing ends up \emph{evolving} backwards, 
 an ignored backward directed field will result
 in miscalculated nonlinear effects, 
 since the total field $\Vec{E} = \Vec{E}^+ + \Vec{E}^-$ 
 will be different to the assumed value of $\Vec{E}^+$.

\emph{No accumulation} 
 occurs when the evolution of any backward directed field $\Vec{E}^-$
 is dominated by its coupling via the residual terms
 to the forward directed field $\Vec{E}^+$; 
 and not by its preferred underlying backward evolution.
No accumulation means that forward evolving field components
 do not couple to field components that evolve backward; 
 this the typical behaviour since the phase mismatch 
 between forward evolving and backward evolving components
 is $\sim 2\beta$; 
 in essence it is comparable
 to the common rotating wave approximation (RWA).
This rapid relative oscillation
 means that backward evolving components never accumulate, 
 as each new addition will be out of phase with the previous one; 
 it is not quite a ``no reflection'' approximation, 
 but one that asserts that the many micro-reflections
 will not combine to produce something significant.
An estimate of the conditions required to break this approximation
 are given in Appendix \ref{S-RWA}; 
 generally speaking this is a much more robust approximation
 than the slow evolution one.
Of course, 
 periodic spatial modulation of the medium gives periodic residual terms, 
 and these can be engineered to force phase matching.
In most contexts this would be a periodicity based on a relatively 
 small phase mismatch 
 (see e.g. quasi phase matching in Boyd \cite{Boyd-NLO}); 
 but might even go as far as matching the backward wave 
 (see e.g. \cite{Harris-1966apl}).

It is also important to note that the same small size of perturbation
 from the residual terms \emph{can} accumulate on the 
 forward evolving field components
 (or, indeed, 
  the backward perturbation on the backward evolving field components).
Although the magnitude of the residual terms acting on
 the forward and backward field evolution are identical, 
 forward evolving components of the residuals 
 can accumulate on the forward evolving field
 because they are phase matched;  
 whereas backward residuals are not, 
 and rapidly average to zero.

%
% ----------------------------------------------------------------------
\subsection{Polarization and Magnetization}

To see most clearly how different optical effects satisfy
 this slow evolution criteria, 
 I will split the total polarization $\Vec{P}_\epsilon$ 
 into pieces:
~
\begin{align}
  \mu_L(t) 
  \convol
  \Vec{P}_\epsilon(\Vec{E},\Vec{r},t)
&=
  \phi_\epsilon (\Vec{E},t)
   \convol
  \Vec{E}(\Vec{r},t)
 +
  \Vec{V}_\epsilon (\Vec{E},\Vec{r},t)
\nonumber
\\
&=
  \phi_L (\Vec{E},t)
   \convol
  \Vec{E}(\Vec{r},t)
\nonumber
\\
&\quad~
 +
  \phi_N (\Vec{E},t)
   \convol
  \Vec{E}(\Vec{r},t)
\nonumber
\\
&\qquad
 +
  \Vec{V}_L (\Vec{E},\Vec{r},t)
 +
  \Vec{V}_N (\Vec{E},\Vec{r},t)
.
\end{align}
The part which is scalar in nature is represented by $\phi_\epsilon$, 
 it might contain linear parts and time response ($\phi_L$); 
 but can also be a function of transverse wave vector
 (i.e. be angle dependent), 
 or contain nonlinear contributions $\phi_N$ such as the 
 third order Kerr nonlinearity
 with $\phi_N \Vec{E} \propto (\Vec{E} \cdot \Vec{E}) \Vec{E}$.
The vector part $\Vec{V}_\epsilon$ would typically be e.g. 
 a second order nonlinearity, 
 which couples the ordinary and extra-ordinary field polarizations.
Note that this description of the material parameters 
 does not restrict allowed values
 of $\epsilon$ in any way; 
 they can include any order of nonlinearity.

The same can be done for $\Vec{M}_\mu$, 
 the non-isotropic and nonlinear
 (i.e. the non-$\mu_L$)
 part of the magnetization.
However, 
 the calculations will all follow the same basic pattern
 that they do for $\Vec{P}_\epsilon$, 
 albeit somewhat complicated by the curl operation.
Since magnetic nonlinearity is rarely present when 
 considering optical propagation, 
 I leave detailed assessment of such effects to later work.

%
% ----------------------------------------------------------------------
\subsection{Residual terms and slow evolution}

Now I will treat each possible residual term in order, 
 where the oppositely directed field is negligible:
 i.e., for $\Vec{E}^\pm$, we have that $\Vec{E}^\mp\simeq 0$.
 where the scalar $\epsilon_L$ contains the
 linear response of the material that is both isotropic 
 and lossless (or gain-less); 
 since here it is a time-response function, 
 it is convolved with the electric field $\Vec{E}$.
Note that the field vectors $\Vec{E}, \Vec{D}$, 
 and indeed the material parameter $\epsilon_L$
 are all functions of time $t$ and space $\Vec{r} = (x,y,z)$; 
 the polarization $\Vec{P}_\epsilon$ and its components
 $\phi_\epsilon$, $\Vec{V}_\epsilon$
 are a functions of time $t$, space $\Vec{r}$, 
 and the field $\Vec{E}$.

Below I will refer to field components ${E}_i$, 
 where $\Vec{E} \equiv ({E}_x,{E}_y,{E}_z)$ and $i \in \{x,y,z\}$; 
 also to wave vector components $k_i$ from $\Vec{k} = (k_x, k_y, k_z)$, 
 with $k_\perp^2 = k_x^2 + k_y^2$.
However, 
 note that in the constraints below, 
 that $k_\perp$ is also used as a substitute symbol
 to represent any one of $k_x$, $k_y$, or $k_\perp$.

~\\
\emph{Firstly,} 
 we have the diffraction term
 $\grad_\perp^2 \Vec{E}$, 
 which is linear.
%I write the vector components of $\Vec{E}$ as ${E}_x, {E}_y, {E}_z$, 
% and denote a single component ${E}_i$.
For $i, j \in \{ x, y \}$, 
 and in transverse wave vector space, 
 the criteria is
~
\begin{align}
  \frac{\imath k_j^2 \left| {E}_i^+ + {E}_i^- \right| / 2 \beta}
       {\imath \beta \left| {E}_i^\pm \right|}
&\simeq
  \frac{\imath k_j^2 \left| {E}_i^\pm \right|/ 2 \beta}
       {\imath \beta \left| {E}_i^\pm \right| }
&=
  \frac{k_j^2}
       {2 \beta^2}
%\quad
\ll
  1
.
\label{eqn-uni-diffraction}
\end{align}
This is just the criterion already given in 
 \cite{Kinsler-2008-fbdiff}, 
 and is identical to the standard paraxial criteria.
This diffraction constraint
 applies only to the transverse behaviour of the pulse, 
 it does not constrain the pulse's intensity, 
 temporal bandwidth, 
 or field profile in any way.

~\\
\emph{Secondly,}
 scalar polarization terms $\phi_\epsilon$,
 which can be either linear ($\phi_L$) 
 or nonlinear ($\phi_N(\Vec{E})$).
These might encode e.g. some of the dispersion, 
 birefringence, 
 or perhaps an angle-dependent refractive index; 
 if nonlinear they might arise from e.g. a third-order nonlinearity.
Such terms give us the criterion
~
\begin{align}
  \frac{\imath \phi_\epsilon \left| {E}_i^++{E}_i^- \right|/ 2 \beta}
       {\imath \beta \left| {E}_i^\pm \right| }
&\simeq
  \frac{\imath \phi_\epsilon \left| {E}_i^\pm \right| / 2 \beta}
       {\imath \beta \left| {E}_i^\pm \right| }
&=
  \frac{\phi_\epsilon}
       {2 \beta^2}
\quad
\ll
  1
.
\label{eqn-uni-scalarapprox}
\end{align}
In the linear case, 
 $\phi \equiv \phi_L$ is independent of $\Vec{E}$, 
 so only the material parameters are constrained, 
 the pulse properties play no role.
In the nonlinear case, 
 e.g. for a third-order nonlinearity, 
 as already treated in \cite{Genty-KKD-2007oe,Kinsler-2007josab},
 we have $\phi \equiv \phi_N \simeq \chi^{(3)} | \Vec{E}^+ |^2$.
Thus the nonlinear criteria makes demands
 on the peak intensity of the pulse --
 but does not apply smoothness assumptions or bandwidth restrictions.

~\\
\emph{Thirdly,}
 linear and nonlinear terms from $\Vec{V}_\epsilon$.
These will have a criterion broadly the same
 as the scalar cases in eqn. \eqref{eqn-uni-scalarapprox},
 but with $\Vec{V}_\epsilon$ replacing $\phi_\epsilon \Vec{E}$.
Thus for $i \in \{x,y,z\}$, 
 we can write down constraints for each component
 of the vector $\Vec{V}_\epsilon$,
 which are
~
\begin{align}
  \imath k_0^2 \left| {V}_{\epsilon,i} \right| / 2 \beta
& \ll
  \imath \beta \left| {E}_i^\pm \right|
& \Longrightarrow \quad
  \left| {V}_{\epsilon,i} \right|
& \ll
  2 \frac{\beta^2}{k_0^2} \left| {E}_i^\pm \right|
.
\label{eqn-uni-vectorapprox}
\end{align}
In the linear case, 
 $\Vec{V}_\epsilon \equiv \Vec{V}_L$, 
 and since $\Vec{V}_L$ and $\Vec{E}$ have some linear relationship, 
 this criterion only constrains the material parameters
 contained in $\Vec{V}_L$,
 not the pulse.
In the nonlinear case, 
 $\Vec{V}_\epsilon \equiv \Vec{V}_N$, 
 the same holds except just as for scalar nonlinear terms, 
 the peak pulse intensity is restricted; 
 e.g. for a $\chi^{(2)}$ medium, 
 $|\Vec{V}_N| \sim \chi^{(2)} |\Vec{E}|$.

However, 
 one complication of the vector cases is that a field consisting
 of only one field polarization component (e.g. ${E}_x^+$)
 may induce a driving in the orthogonal 
 (and initially zero) components (e.g. ${E}_y^\pm$).
Hence
 both ${E}_y^\pm$ fields will be driven with the same strength, 
 so that it is far from obvious that we can set ${E}_y^-$ to zero, 
 but still keep the ${E}_y^+$ without being inconsistent.
However, 
 as described above, 
 it is the phase matching which ensures that forward residuals accumulate, 
 whilst the non-matched backward residuals are subject to the RWA, 
 and become negligible:
 hence we can still rely on eqn. \eqref{eqn-uni-vectorapprox}, 
 albeit under caution.

~\\
\emph{Fourthly,}
 we have the divergence term $\grad \grad \cdot \Vec{P}_\epsilon$.
Often this term is considered negligible, 
 and discarded even before writing down the second order wave equation; 
 nevertheless we should test it.
Here we consider just scalar linear or nonlinear terms $\phi$, 
 but the arguments can be adapted to vector terms 
 as done above; 
 in any case the results are comparable.
For $i,j \in \{ x, y, z \}$, 
 we have
~
\begin{align}
  \imath k_i k_j \left| P_j^++P_j^- \right| / 2 \beta
& \ll
   \imath \beta \left| {E}_i^\pm \right|
\nonumber
\\
  \frac{k_i k_j}{2\beta^2} \phi 
   \left| {E}_j^++{E}_j^- \right|
& \ll
   \left| {E}_i^\pm \right|
.
\label{eqn-uni-divergence}
\end{align}
There are four distinct cases to consider here, 
 but only two resulting criteria.
First, 
 if $i \in \{x,y\}$, 
 then whether $j \in \{x,y\}$ or $j \equiv z$
 we find that 
~
\begin{align}
  \frac{k_\perp^2}{2\beta^2} \left| \phi \right|
&\ll
  1
\label{eqn-uni-divergence-xy}
\end{align}
since $|{E}_z|/|{E}_i| \sim k_\perp/\beta$; 
 this we see that this is a combination of both the diffraction
 and nonlinear criteria, 
 and is thus easily satisfied.
For the second, 
 where $i \equiv z$, 
 all the wave vector contributions cancel,
 leaving simply
~
\begin{align}
  \left| \phi \right|
&\ll
  1
.
\label{eqn-uni-divergence-z}
\end{align}
It is thus directly comparable to the scalar nonlinear criteria above, 
 and equally likely to be satisfied; 
 the comparable vector criteria are $k_\perp^2 {V}_i / 2\beta^2 \ll {E}_i$
 and ${V}_i \ll {E}_i$.

~\\
\emph{Fifthly,}
 we must consider the charge density $\rho$ and charge current $\Vec{J}$.
These criteria are simple to write down, 
 but whether they are satisfied will depend
 on the initial conditions and the response of how these are modeled
 to the propagating pulse.
This is something that may need to be checked during simulation
 or solution of the pulse propagation, 
 and not assumed beforehand, 
 although Berge and Skupin \cite{Berge-S-2009dcdsa}
 discuss the issues in the context of optical beam filamentation.
The charge and current constraints are
~
\begin{align}
  \frac{\left| \rho \right|}
       {2 \beta^2 \epsilon_0 \left| \epsilon_L \right|}
&\ll
  \left| {E}_i \right|
,
\label{eqn-uni-charge}
\\
  \frac{c k_0 \mu_0 \mu_L \left| {J}_i \right|}
       {2 \beta^2}
&\ll
  \left| {E}_i \right|
.
\label{eqn-uni-current}
\end{align}

~\\
\emph{Sixthly,}
 a constraint on the non-$\mu_L$ magnetization $\Vec{M}_\mu$
 can also be written down, 
 although (as already discussed) I leave the details
 for later work.
It is
~
\begin{align}
  \frac{c k_0 \mu_0}
       {2 \beta^2}
  \left|
    \grad \cross \vec{M}_\mu
  \right|
&\ll
  \left| E_i \right|
.
\label{eqn-uni-curlM}
\end{align}
Here the curl operator might often be expected to return
 a value of order $\beta$, 
 so with $k_0 \sim \beta$ we have
 $c|\Vec{M}_\mu|/2 \ll |E_i|$.

~\\
\emph{To summarize}, 
 the diffraction criterion asserts the beam must be sufficiently paraxial, 
 the linear criteria asserts the material must have weak dispersion, 
 and the nonlinear criteria assert the nonlinear effect must be weak.
Paraxiality is determined by our experimental conditions, 
 and can thus be guaranteed if desired, 
 and for most optical materials, 
 the dispersion is sufficiently weak -- 
 except perhaps in the vicinity of resonances or band gaps.
Weak nonlinearity is invariably guaranteed by material damage thresholds, 
 since the material suffers damage long before nonlinear effects 
 become strong -- 
 nevertheless, 
 the effects of such strong nonlinearities
 on uni-directional approximations
 have been analytically and numerically studied \cite{Kinsler-2007josab}.
Finally, 
 it is worth noting that each criterion is independent of the others, 
 so each effect can be tested for separately.

%
% ----------------------------------------------------------------------
\subsection{Uni-directional equation for $\Vec{E}^+$}

In the case where all of the wavelength-scale 
 slow-evolution criteria listed above hold, 
 we can be sure that the backward directed field $\Vec{E}^-$ is negligible, 
 and if the no-accumulation condition also holds, 
 then neither will there be any backward evolving contributions to the field.
Consequently, 
 we can be sure that an initially negligible $\Vec{E}^-$ remains so, 
 and again with $k_0 = \omega / c$, 
 the bi-directional eqn. \eqref{eqn-bi-dzE} simplifies to 
~
\begin{align}
    \partial_z
  \Vec{E}^+(\omega)
&=
 +
    \imath
    \beta(\omega)
  \Vec{E}^+(\omega)
 \quad
 +
  \frac{\imath \grad_\perp^2}{2 \beta(\omega)}
    \Vec{E}^+(\omega)
\nonumber
\\
&\quad
 +
  \frac{\imath k_0^2(\omega) \mu_L}
       {2 \beta(\omega)}
  \Vec{P}_\epsilon
  (
    \Vec{E}^+, \omega
  )
 -
  \frac{c k_0(\omega) \mu_0 \mu_L}
       {2 \beta(\omega)}
  \Vec{J}(\omega)
\nonumber
\\
&\quad ~~
 -
  \frac{c k_0(\omega) \mu_0}
       {2\beta(\omega)}
  \grad \cross
  \Vec{M}_\mu(\Vec{H}^+)
\nonumber
\\
& \qquad
 +
  \frac{\imath}
       {2 \beta(\omega) \epsilon_L(\omega)}
  \grad
  \left[
    \grad \cdot \Vec{P}_\epsilon(\Vec{E}^+, \omega)
   -
    \frac{\rho(\omega)}
         {\epsilon_0}
  \right]
.
\label{eqn-uni-dzE}
\end{align}
Here now the polarization $\Vec{P}_\epsilon$, 
 diffraction, 
 and divergence are solely dependent 
 on the forward directed field ($\Vec{E}^+$).
Likewise the magnetization term $\Vec{M}_\mu$ 
 should be considered as being solely dependent
  on the forward directed field ($\Vec{H}^+$) -- 
 although we will need to estimate the value of $\Vec{H}^+$
 using the known electric field $\Vec{E}^+$.
Since we are in a slow evolution approximation, 
 a good estimate for the components of $\Vec{H}^+$ 
 will simply be those of $\Vec{E}^+$ scaled by 
 $\epsilon_0 (\epsilon_L/\mu_L)^{1/2} c$; 
 so that (e.g.) $H_y^+$ depends on $E_x^+$.
Also, 
 the $\grad \cross \Vec{M}_\mu$ will be dominated
 by the $z$ dependence of its $x$ and $y$ components, 
 so that it will typically generate factors
 of order $\beta |\Vec{M}_\mu|$.

Although I have included magnetic effects 
 in the derivation of eqn. \eqref{eqn-uni-dzE}, 
 I do not consider specific cases in detail, 
 as has been done for plane-polarized light in e.g. 
 \cite{Scalora-SAPDMBZ-2005prl,Wen-XDTSF-2007pra,DAguanno-MB-2008josab}.
The derivations in those articles are ``traditional'' 
 in the sense that each consists of multiple interim stages 
 at which an additional approximation is applied; 
 it is instructive to compare those derivations with mine.
In particular, 
 e.g. all apply bandwidth limitations, 
 and discard various high-order derivative terms 
 that are not specific to their choice of propagation medium.
Although Scalora et al. \cite{Scalora-SAPDMBZ-2005prl}
 is the least aggressive in this respect, 
 it does not allow for magnetic nonlinearity.

%
% ======================================================================
\section{Modifications}
\label{S-modifications}

Let us now consider some of the strategies used in other approaches,
 some of which were \emph{required} in order to get approximations 
 that eventually achieved a sufficiently simple evolution equation.
In particular, 
 the various envelope equations 
 (e.g. \cite{Brabec-K-1997prl,Kinsler-N-2003pra,Geissler-TSSKB-1999prl,
         Scalora-C-1994oc}, 
   and even \cite{Wen-XDTSF-2007pra,DAguanno-MB-2008josab})
 all use co-moving frames and/or envelopes
 as a preparation for discarding inconvenient derivatives:
 here such steps are optional extras.
In this factorization approach shown here, 
 none of these were required, 
 but they nevertheless may be useful.
Examples are as follows:

\begin{enumerate}

\item
A \emph{co-moving frame} can now be added, 
 using $t'=t-z/v_f$.
This is a simple linear process that causes no extra complications;
 the leading RHS $\imath \beta {E}^+$ term is replaced
 by $\imath (\beta \mp k_f) {E}^\pm$,
 for frame speed $v_f = \omega_1/k_f$.
Note that setting $\beta=k_f$ will freeze the phase velocity
 of a pulse centred at $\omega_1$, 
 not the group velocity.

\item
The \emph{field can be split up} into pieces localized
 at certain frequencies, 
 as done in descriptions of OPAs or Raman combs
 (as in e.g. \cite{Kinsler-N-2003pra,Kinsler-N-2004pra,Kinsler-N-2005pra}).
 The wave equation can then be separated into one equation for 
 each piece, 
 coupled by the appropriate frequency-matched polarization terms
 (see e.g. \cite{Kinsler-NT-2006-phnlo}).

\item
A \emph{carrier-envelope} description of the field is not required, 
 but can easily be implemented
 with the usual prescription of \cite{Gabor-1946jiee,Boyd-NLO}
 ${E}(t) = A(t) \exp [\imath (\omega_1 t - k_1 z)] 
        + A^*(t) \exp [-\imath (\omega_1 t - k_1 z)] $
 defining the envelope $A(t)$ 
 with respect to carrier frequency $\omega_1$ and wave vector $k_1$;
 this also provides a built in a co-moving frame
 $v_f = \omega_1/k_1$.
Multiple envelopes centred at different carrier frequencies
 and wave vectors ($\omega_i$, $k_i$) can also be used
 \cite{Kinsler-NT-2006-phnlo,Boyd-NLO}.

\item
\emph{Bandwidth restrictions} might be added (see below), 
 either to ensure
 a smooth envelope or to simplify the wave equations; 
 in addition they might be used to separate out or neglect 
 frequency mixing terms or harmonic generation.
As it stands, 
 no bandwidth restrictions were applied when deriving 
 eqn. \eqref{eqn-uni-dzE} -- 
 there are only the limitations of the dispersion and/or polarization models
 to consider.

\item
\emph{Mode averaging} is where the transverse extent
 of a propagating beam is not explicitly modeled, 
 but is subsumed into a description of a transverse mode profile; 
 as such it is typically applied to situations involving 
 optical fibres or other waveguides.
See e.g. \cite{Laegsgaard-2007oe} for a recent approach, 
 which goes beyond a simple addition of a frequency dependence
 to the ``effective area'' of the mode, 
 and generalizes the effective area concept itself.

\end{enumerate}

A wave equation like that derived above,
 but limited to describing propagation in optical fibres 
 (i.e. a dispersive and third order nonlinear material),
 has already been studied \cite{Genty-KKD-2007oe}; 
 but it did not consider the effects of diffraction
 or angle dependent refractive index, 
 vector polarization terms, 
 or the divergence of $\Vec{P}_\epsilon$.
It did, 
 however, 
 show a stunning level of agreement between 
 uni-directional envelope and PSSD \cite{Tyrrell-KN-2005jmo} Maxwell equations 
 simulations
 in the case of optical carrier wave shocking -- 
 even though it described the pulse using an envelope!

If desired, 
 we can easily recover wave equations that match
 the SEWA and SVEA wave equations already in common use, 
 by applying bandwidth constraints to our field, 
 and making approximations based on them.
First, 
 we set $k_0 = \omega_0 ( 1 + \delta ) / c$, 
 with $\delta = (\omega - \omega_0) / \omega_0$.
Then assume that our field $\Vec{E}^+$ has a bandwidth 
 much smaller than the carrier frequency $\omega_0$, 
 so that $\Vec{E}(\omega_0 ( 1 + \delta ))$ is only non-negligible
 for $\delta \ll 1$; 
 thus we can now assume $\delta^2 \simeq 0$.
This bandwidth constraint amounts to an assumption about the smoothness
 of the pulse in the time domain.
The $k_0^2$ factor now simplifies
 to $k_0^2 \simeq \omega_0^2 (1 + 2 \delta)/c^2$, 
 and hence we get a \emph{non-envelope} but otherwise
 SEWA-like wave equation \cite{Brabec-K-1997prl},
 which is
~
\begin{align}
    \partial_z
  \Vec{E}^+(\omega)
&=
 +
    \imath
    \left( \beta(\omega) - k_f \right)
  \Vec{E}^+(\omega)
 ~~
 +
  \frac{\imath}{2 \beta(\omega)}
  \grad_\perp^2
    \Vec{E}^+(\omega)
\nonumber
\\
&\quad
 +
  \frac{\imath \omega_0^2 \mu_L}
       {2 c^2 \beta(\omega)}
  \left[
    1 + 2 \frac{\omega-\omega_0}{\omega_0}
  \right]
  \Vec{P}_\epsilon(\Vec{E}^+(\omega),\omega)
.
\label{eqn-uni-dzE-SEWA}
\end{align}
The next level of bandwidth-limiting approximation
 takes us back to an equation matching the venerable SVEA.
To achieve this
 we take such narrow-band fields that we can set $\delta \simeq 0$, 
 and so 
~
\begin{align}
    \partial_z
  \Vec{E}^+(\omega)
&=
 +
    \imath
    \left( \beta(\omega) - k_f \right)
  \Vec{E}^+(\omega)
 ~~
 +
  \frac{\imath}{2 \beta(\omega)}
  \grad_\perp^2
    \Vec{E}^+(\omega)
\nonumber
\\
&\quad
 +
  \frac{\imath \omega_0^2 \mu_L}
       {2 c^2 \beta(\omega)}
  \Vec{P}_\epsilon(\Vec{E}^+(\omega),\omega)
.
\label{eqn-uni-dzE-SVEA}
\end{align}
Neither of these (SEWA-like, SVEA-like) wave equations 
 are required to incorporate an envelope-carrier description of the fields, 
 or a co-moving frame
 as demanded by the usual SEWA or SVEA derivations; 
 the moving frame specified by $k_f$ above is a mere convenience, 
 and $k_f$ may be set to zero.
Strictly speaking, 
 to match the SEWA or SVEA wave equations most closely, 
 we should also set $\beta$ to a fixed value, 
 and put \emph{all} of the remaining linear dielectric properties
 of the material
 into $\Vec{P}_\epsilon$.

Even in the SVEA limit, 
 the factorization technique allows us to 
 recover the same propagation equations 
 as derived using standard approaches, 
 but this derivation now gives us a better (and much simpler)
 basis on which to judge their robustness
 to strong nonlinearity, 
 angle dependent refractive indices, 
 and diffraction or transverse effects.
Note in particular that the linear constraints given
 in section \ref{S-firstorder} depend \emph{only} on the 
 material properties, 
 and not on the field in any way.
The nonlinear constraints are the same, 
 but with an additional dependence on the peak field strength --
 but importantly, 
 \emph{not} its smoothness or bandwidth.

It is important to remember that 
 introducing an envelope and carrier representation of the pulse
 remains useful. %,  for at least two reasons.
This is because %Firstly, 
 a well chosen carrier frequency $\omega_1$ 
 will almost certainly provide
 an envelope smoother than the field itself; 
 this will provide a more intuitive picture
 but will also have advantages for numerical computation.

%
% ======================================================================
\section{Examples}
\label{S-examples}

%
% ----------------------------------------------------------------------
\subsection{Third order nonlinearity}

Third order nonlinearities are common in many materials, 
 e.g. in the silica used to make optical fibres \cite{Agrawal-NFO}.
Here I study propagation in a comparable material,
 but also allow for magnetic dispersion.
The propagation is 
 based around a wave vector reference $\beta$, 
 where the residual frequency dependence of the material refractive index
 is represented by a dimensionless parameter $\kappa$
 dependent on the linear dispersive parts
 of the permittivity $\epsilon_d$
 and permeability $\mu_d$, 
 so that 
 $\kappa = \omega (\epsilon_d \mu_d)^{1/2}/\beta - 1$.
The instantaneous electric third order nonlinearity
 is $\chi^{(3)} $.
For plane polarized fields, 
 the uni-directional wave equation for ${E}_x^{+}(\omega)$
 can be derived from eqn. \eqref{eqn-uni-dzE}, 
 and with the usual $k_0=\omega/c$ is
~
\begin{align}
  \partial_z {E}_x^{+}
&=
 +
  \imath 
    \beta
  \left[
    1
   +
    \kappa
  \right]
  {E}_x^{+}
 ~
 +
  \frac{\imath k_0^2 \mu_L}
       {2 \beta}
  \mathscr{F}
  \left[
    \chi^{(3)}
    {E}_x^2(t)
    {E}_x^{+}(t)
  \right]
\nonumber
\\
& \qquad
 +
  \frac{\imath \grad_\perp^2}
       {2 \beta}
    {E}_x^+
\label{eqn-eg-Enls}
,
\end{align}
 where $\mathscr{F}[...]$ is the Fourier transform that 
 converts the time-domain nonlinear polarization 
 into its frequency domain form.

This is a generalized nonlinear Schr\"odinger (NLS) equation, 
 but is for the full field (i.e. uses no envelope description)
 and retains the full nonlinearity 
 (i.e. retains the third harmonic generation term).
The only assumptions made are that of
 transverse fields, 
 weak dispersive corrections $\kappa$, 
 and weakly nonlinear response; 
 these all allow us to decouple
 the forward and backward wave equations.
This decoupling then allows us, 
 without any extra approximation,
 to reduce our description to one of forward only pulse propagation.
The specific example chosen here
 is for an instantaneous cubic nonlinearity, 
 but it is easily generalized to non-instantaneous cases
 or other scalar nonlinearities.

We can transform eqn. \eqref{eqn-eg-Enls}
 into a NLS equation by 
 representing the field in terms of an envelope and carrier, 
 where the carrier has a fixed frequency $\omega_1$ and 
 wavevector $k_1$; 
 i.e. using
~
\begin{align}
  {E}_x^+(t) 
&=
  A(t) \exp \left[\imath \left(\omega_1 t - k_1 z\right)\right]
\nonumber
\\
& \qquad
        + A^*(t) \exp \left[-\imath \left(\omega_1 t - k_1 z\right)\right] 
.
\label{eqn-eg-envcar}
\end{align}
In the frequency domain an arbitrary frequency 
 $\omega$ differs from the carrier frequency $\omega_1$
 by an offset $\Delta$; 
 i.e. $\omega = \omega_1 + \Delta$; 
 hence the frequency domain counterpart to $A(t)$ 
 is best written $A(\Delta)$, 
 not $A(\omega)$.
We proceed by setting $\beta$ to have the constant value $k_1$,  
 and ignoring the off-resonant THG term, 
 which is usually very poorly phase matched.
After separating into a pair of complex-conjugate equations
 (one for $A$, one for $A^*$),
 this gives us the expected NLS equation
 with diffraction.
The chosen carrier effectively moves us into a frame
 that freezes those carrier oscillations, 
 but this differs from one that is co-moving with the pulse envelope, 
 i.e. one moving at the group velocity $v_g=\partial\omega/\partial k$.
After we transform into a frame co-moving with the group velocity, 
 where at $\omega_1$ we have $K_g=\omega_1 (v_g^{-1}-v_p^{-1})$,
 the frequency domain wave equation is
~
\begin{align}
  \partial_z {A}
&=
 +
  \imath 
  K(\Delta)
  {A}
 ~
 +
  \frac{\imath k_0^2 \mu_L}
       {2 k}
    \mathscr{F}
    \left[
      \chi^{(3)}
      {\left| A(t) \right|^2}
      {A}(t)
    \right]
%\nonumber
%\\
%& \qquad
 +
  \frac{\imath \grad_\perp^2}
       {2 k}
    {A}
,
\label{eqn-eg-nls}
\end{align}
 with $K(\Delta) = k \kappa(\omega_1+\Delta) + K_g$.
All that has been assumed to derive this equation
 is uni-directional propagation 
 and negligible third harmonic generation.
This eqn. \eqref{eqn-eg-nls} is for a magnetically dispersive system
 broadly comparable to that giving rise to the eqn. (12) 
 of Scalora et al. \cite{Scalora-SAPDMBZ-2005prl}
 (henceforth eqn. (S12)\footnote{Note that eqn. (S12)
  has scaled both the time and space parameters.}); 
 although I have additionally retained diffraction
 and any order of dispersion.

Many instances of NLS-type equations, 
 such as that of eqn. (S12) or simpler forms (e.g. \cite{Agrawal-NFO}),
 are written in the time domain, 
 which means that it is more complicated to represent
 the full range of the dispersive response.
When transforming eqn. \eqref{eqn-eg-nls} into the time domain, 
 the dispersion term $K(\Delta)A(\Delta)$ becomes a convolution --  
 but it can also be represented as a Taylor series in time derivatives.
This Taylor series is usually reduced to a few low order terms, 
 and when using the correct group velocity, 
 the lowest order term is a quadratic.
Also often seen in NLS equations 
 is the self-steepening term 
 (again see eqn. (S12)).
This self-steepening term be obtained from eqn. \eqref{eqn-eg-nls} 
 by expanding $k_0^2=\omega^2/c^2=(\omega_1+\Delta)^2/c^2$, 
 in a similar manner to deriving a SEWA-like equation
 as discussed in the previous section.
Then the leading term ($\propto \omega_1^2$)
 gives the usual nonlinear term, 
 whilst the first order %(in $\Delta$) 
 contribution ($\propto 2\omega_1\Delta$)
 gives the single time derivative
 needed for self-steepening in the time domain.
Also present in eqn. (S12),
 but \emph{not} in eqn. \eqref{eqn-eg-nls}
 is a term proportional to ${\chi^{(3)}}$ squared.
Here such a term is not present because it is second order correction, 
 whilst the uni-directional approximation applied here is first order.
Whilst it is possible to incorporate higher-order corrections, 
 one has to be careful to remain consistent, 
 and not miss other significant corrections of the same order, 
 nor to include unnecessary terms
 which should strictly be considered negligible.

%
% ----------------------------------------------------------------------
\subsection{Second order nonlinearity}

The case of second order nonlinearity 
 is a little more complicated, 
 since it typically couples the two possible polarization states
 of the field together \cite{Boyd-NLO}.
For simplicity, 
 I will avoid an exhaustive, 
 detailed derivation from first principles, 
 and instead just give example wave equations directly.
Indeed,
 they can be easily inferred directly from the format of the coupling
 in standard treatments.

In second-order nonlinear interactions
 such as optical parametric amplification (OPA)
 in lithium borate (LBO) using birefringent phase-matching, 
 two field polarizations need to be considered. 
To model the cross-coupling between the orthogonally-polarized fields,
 it is necessary to solve for both field polarizations; 
 and to allow for the birefringence we need a pair of 
 linear responses, 
 i.e. $\kappa_{x}(\omega), \kappa_{y}(\omega)$.

Since it is convenient, 
 I split the vector form of the $\Vec{E}^\pm$ wave equation 
 up into its transverse $x$ and $y$ components.
The propagation is 
 based around a wave vector reference $\beta$, 
 where the residual frequency dependence of the material refractive index
 in the $x$ or $y$ directions 
 is represented by a dimensionless parameter $\kappa_i$,
 for $i \in \{x,y\}$.
This $\kappa_i$ is
 dependent on the linear dispersive parts 
 of the permittivities $\epsilon_{d,i}$
 and permeabilities $\mu_{d,i}$, 
 so that 
 $\kappa_i = \omega (\epsilon_{d,i} \mu_{d,i})^{1/2}/\beta - 1$.
The instantaneous electric second order nonlinear coefficient
 is $\chi^{(2)}$.
Based on eqn. \eqref{eqn-uni-dzE}, 
 and for second harmonic generation in the orthogonal polarization
 (i.e. a type I OPA), 
 the wave equations for ${E}_x^{+}(\omega)$ and ${E}_y^{+}(\omega)$
 (with the usual $k_0=\omega/c$)
 are
~
\begin{align}
  \partial_z {E}_x^{+}
&= 
 +
  \imath 
    \beta
  \left[
    1
   +
    \kappa_x
  \right]
  {E}_x^{+}
 ~~
\nonumber
\\
& \quad~~
 +
  \frac{\imath k_0^2 \mu_L}
       {2 \beta}
  \mathscr{F}
  \left[
    2 \chi^{(2)}
    {E}_y^+(t)
    {E}_x^{+}(t)
  \right]
 +
  \frac{\imath \grad_\perp^2}
       {2 \beta}
    {E}_x^+
\label{eqn-chi2-x}
\\
  \partial_z {E}_y^{+}
&= 
 +
  \imath 
    \beta
  \left[
    1
   +
    \kappa_y
  \right]
  {E}_y^{+}
 ~~
\nonumber
\\
& \quad~~
 +
  \frac{\imath k_0^2 \mu_L}
       {2 \beta}
  \mathscr{F}
  \left[
     \chi^{(2)}
    {E}_x^{+2}(t)
  \right]
 +
    \frac{\imath \grad_\perp^2}{2 \beta}
    {E}_y^+
,
\label{eqn-chi2-y}
\end{align}
 where $\mathscr{F}[...]$ is the Fourier transform that 
 converts the time-domain nonlinear polarization 
 into its frequency domain form.
The specific example chosen here
 is easy to modify to allow for or incorporate other $\chi^{(2)}$ processes.
Remarkably, 
 it is also strikingly similar in appearance
 (although not in detail) 
 to the usual SVEA equations used to propagate narrow-band pulse envelopes; 
 despite the lack of a co-moving frame, 
 and even though they are for the field, 
 not an envelope.

We can transform eqns. \eqref{eqn-chi2-x}, \eqref{eqn-chi2-y}
 into the usual equations for a parametric amplifier
 by representing the $x$ and $y$ polarized fields 
 in terms of three envelopes and carrier pairs:
~
\begin{align}
  {E}_x(t) 
&= 
  A_1(t) \exp \left[\imath \left(\omega_1 t - k_1 z\right)\right]
\nonumber
\\
&\qquad
        + A_1^*(t) \exp \left[-\imath \left(\omega_1 t - k_1 z\right)\right]
\nonumber
\\
& \qquad\quad
 +
  A_2(t) \exp \left[\imath \left(\omega_2 t - k_2 z\right)\right]
\nonumber
\\
& \qquad\qquad
        + A_2^*(t) \exp \left[-\imath \left(\omega_2 t - k_2 z\right)\right]
\label{eqn-eg-envcar12}
\\
  {E}_y(t) 
&= 
  A_3(t) \exp \left[\imath \left(\omega_3 t - k_3 z\right)\right]
\nonumber
\\
&\qquad
        + A_3^*(t) \exp \left[-\imath \left(\omega_3 t - k_3 z\right)\right] 
\label{eqn-eg-envcar3}
\end{align}
 where $\omega_3 = \omega_1 + \omega_2$.
After separating into pairs of complex-conjugate equations
 (one each for $A_i$, one for $A_i^*$), 
 and ignoring the off-resonant polarization terms,
Just as for the NLS example above, 
 we also transform into a frame co-moving with the group velocity,
 although here we select the group velocity of a preferred frequency component
 (perhaps $\omega_3$),
 with e.g. $K_{g}=\omega_3 (v_{g}^{-1}-v_p^{-1})$.
Choosing $\beta$ for each equation differently, 
 i.e. with $\beta \in \{k_1, k_2, k_3\}$,
 the wave equations for the ${A}_i(\omega)$ are
~
\begin{align}
  \partial_z {A}_1
&= 
% +
  \imath 
    K_1(\Delta)
  {A}_1
 ~
\nonumber
\\
& \quad
 +
  \frac{\imath k_0^2 \mu_L}
       {2 k_1}
  \mathscr{F}
  \left[
    2 \chi^{(2)}
    {A}_3(t) {A}_2^*(t)
  \right]
    e^{-\imath \Delta k z}
 +
  \frac{\imath \grad_\perp^2}
       {2 k_1}
    {A}_1
\label{eqn-chi2-A1}
\\
  \partial_z {A}_2
&= 
% +
  \imath 
    K_2(\Delta)
  {A}_2
 ~
\nonumber
\\
& \quad
 +
  \frac{\imath k_0^2 \mu_L}
       {2 k_2}
  \mathscr{F}
  \left[
    2 \chi^{(2)}
    {A}_3(t) {A}_1^*(t)
  \right]
    e^{-\imath \Delta k z}
 +
  \frac{\imath \grad_\perp^2}
       {2 k_2}
    {A}_2
\label{eqn-chi2-A2}
\\
  \partial_z {A}_3
&= 
% +
  \imath 
    K_3(\Delta)
  {A}_3
 ~
\nonumber
\\
& \quad
 +
  \frac{\imath k_0^2 \mu_L}
       {2 k_3}
  \mathscr{F}
  \left[
     \chi^{(2)}
    {A}_1(t) {A}_2(t)
  \right]
    e^{+\imath \Delta k z}
 +
    \frac{\imath \grad_\perp^2}{2 k_3}
    {A}_3
.
\label{eqn-chi2-A3}
\end{align}
Here $K_{i}(\Delta) = k_{i} \kappa_x(\omega_{i}+\Delta) + K_g$, 
 with $i \in \{1,2\}$; 
 and $K_{3}(\Delta) = k_{3} \kappa_y(\omega_3+\Delta) + K_g$.
The phase mismatch term is $\Delta k = k_3 - k_2 - k_1$.
The only approximations used to derive these equations
 are uni-directional propagation 
 and negligible off-resonant polarization terms.

%
% ======================================================================
\section{Conclusion}
\label{S-conclude}

I have derived a general first order wave equation
 for uni-directional pulse propagation
 that allows for arbitrary dielectric polarization, 
 diffraction, 
 and free electric charge and currents; 
 even magnetic dispersion and other magnetic responses are allowed.
After factorizing the second order wave equation
 into an exact bi-directional model, 
 it applies the same slow-evolution approximation
 to all non-trivial effects (e.g. nonlinearity, diffraction), 
 and so reduces the propagation equations
 to a first order uni-directional wave equation.
My derivation contrasts 
 with typical approaches, 
 which often rely on a co-moving frame 
 and a sequence of different approximations, 
 such as ad-hoc assumption of negligible second derivatives.
In the appropriate limits, 
 it turns out that many existing derivations
 have given similar but more restricted results to those presented here.
As a result, 
 with minimal adjustment, 
 existing numerical and theoretical models could be adapted to take advantage
 of this sounder theoretical basis,
 more straightforward approximations, 
 and simpler error-term calculations.

The improved ``factorization'' derivation presented here
 allows a term-to-term comparison
 of the exact bi-directional theory
 with its approximate uni-directional counterpart, 
 so that the approximation used
 (and its consequences) 
 is much more easily understood.
This means that pulse propagation models
 in the extreme ultrafast and wide-band limits 
 can be made more robust -- 
 since differences between
 exact bi-directional and approximate uni-directional propagation
 can be straightforwardly computed.

%
%=======================================================================
\acknowledgments

I acknowledge financial support from the
 Engineering and Physical Sciences Research Council
 (EP/E031463/1).\\
(Added in v7, May 2014) I am also grateful to Yanfeng Li of Tianjin University
 for passing on the comments of Fanchao Meng; 
 these clarified the assumed homogeneity of $\epsilon_L$ and $\mu_L$, 
 and noted corrections to the sign and scaling of the $\Vec{J}$, 
  $\grad \cdot \Vec{P}$, 
  and $\grad \times \Vec{M}$ terms
 in eqns. \eqref{eqn-basic-nabla2E},  
  \eqref{eqn-basic-nabla2E-Helmholtz},
  \eqref{eqn-bi-d2zE},
  \eqref{eqn-bi-dzE}; 
 these also lead to updates to 
   eqns. \eqref{eqn-uni-current},
  \eqref{eqn-uni-curlM},
  \eqref{eqn-uni-dzE}.

% \cite{EP-E031463-1,GR-T17267-01}

%
%=======================================================================

%\bibliography{/home/physics/_work/bibtex.bib}

%
% ======================================================================

\appendix

%
% ======================================================================
\section{Factorizing}
\label{S-factorize}

Here I present a simple overview of the mathematics
 of the factorization procedure, 
 since full details can be found in \cite{Ferrando-ZCBM-2005pre}.
In the calculations below, 
 I transform into wave vector space, 
 where the $z$-derivative $\partial_z$ is converted to $\imath k$.
Also, 
 we have that $\beta^2 = n^2 \omega^2 /c^2$,
 and the unspecified residual term is denoted $Q$.
The second order wave equation can then be written
~
\begin{align}
  \left[
    \partial_z^2 
   +
    \beta^2
  \right]
  E
&=
 -Q
\\
  \left[
   -k^2 + \beta^2
  \right]
  E
&=
 -Q
\\
  E
&=
  \frac{1}{k^2 - \beta^2}
  Q
\quad
=
  \frac{1}{\left(k-\beta\right)\left(k+\beta\right)}
\\
&=
  \frac{-1}{2\beta}
  \left[
    \frac{1}{k+\beta}
   -
    \frac{1}{k-\beta}
  \right]
  Q
.
\end{align}
Now $(k-\beta)^{-1}$ is a forward-like (Green's function) propagator 
 for the field, 
 but note that in my terminology, 
 it \emph{evolves} the field.
The complementary backward-like propagator
 is $(k+\beta)^{-1}$.
As already described in the main text, 
 we now write ${E}={E}^++{E}^-$, 
 and split the two sides up to get
~
\begin{align}
  {E}^+
 +
  {E}^-
&=
  \frac{-1}{2\beta}
  \left[
    \frac{1}{k+\beta}
   -
    \frac{1}{k-\beta}
  \right]
  Q
\\
  {E}^\pm
&=
  \frac{\pm1}{2\beta}
    \frac{1}{k\mp\beta}
  Q
\\
  \left[
   k \mp \beta
  \right]
  {E}^\pm
&=
 \pm
  \frac{1}{2\beta}
    \frac{1}{k\mp\beta}
  Q
\\
  \imath
  k {E}^\pm
&= 
 \pm
  \imath
  \beta {E}^\pm
 \pm
  \frac{\imath}{2\beta}
  Q
.
\end{align}
Finally, 
 we transform the wave vector space $\imath k$ terms 
 back into normal space to give $z$ derivatives, 
 resulting in the final form
~
\begin{align}
  \partial_z
  {E}^\pm
&= 
 \pm
  \imath
  \beta {E}^\pm
 \pm
  \frac{\imath}{2\beta}
  Q
.
\end{align}

%\newpage

%
% ======================================================================
\section{The no accumulation approximation}
\label{S-RWA}

In the main text, 
 I describe the no accumulation approximation 
 in spectral terms as a RWA approximation.
However, 
 it is hard to set a clear, 
 accurate criterion for the RWA approximation to be satisfied
 in the general case, 
 since it requires knowledge of the entire propagation 
 before it can be justified.
In this appendix, 
 I take a different approach to determine
 the conditions under which the approximation will be satisfied. 

First, 
 consider a forward evolving field
 so ${E} = {E}_0 \exp (\imath k z)$, 
 and therefore 
~
\begin{align}
  {E}_0^-
&=
  \frac{k-\beta}{k+\beta}
  {E}_0^+
\qquad
=
  \xi
  {E}_0^+
,
\end{align}
where as noted $k$ can be difficult to determine, 
 and may even change dynamically; 
 here we can assume it corresponds to the propagation wave vector
 that would be seen at if all the conditions holding at a chosen position
 also held everywhere else.
On this basis, 
 we can even define $k = k(z)$, 
 where by analogy to the linear case we might assert that 
 $k^2(z) = \beta^2 + \mathscr{Q}(z)/E(z)$, 
 so that for small $\mathscr{Q}$, 
 we have
 $k E \simeq \beta ( E + \mathscr{Q} / 2 \beta^2)$.

Let us start by assuming 
 our field is propagating and evolving forwards (only), 
 with perfectly matched ${E}^\pm$ fields; 
 so that ${E}^- = \xi {E}^+$.
 but then it happens that $\mathscr{Q}$ changes by $\delta \mathscr{Q}$
 over a small interval $\delta z$,
 likewise $\xi$ changes by $\delta \xi$.
The ${E}^\pm$ will no longer be matched, 
 and now the total field splits into two parts
 that evolve in opposite directions.
The part that continues to evolve forward has ${E}^+$ nearly unchanged, 
 but the forward evolving ${E}^-$ has changed size 
 (and is now $\propto (\xi - \delta \xi)$)
 to stay perfectly matched according to the new $\mathscr{Q}$.
The rest of the old ${E}^-$ ($\propto \delta \xi$) now propagates backwards, 
 taking with it a tiny fraction of the original ${E}^+$ 
 (and is $\propto \xi \delta \xi$).

Comparing the two backward evolving ${E}^-$ components 
 at $z$ and $z + \delta z$,
 and taking the limit $\delta z \rightarrow 0$
 enables us to estimate that the backward evolving ${E}^-$ field
 changes according to 
~
\begin{align}
  \partial_z
  {E}_{0,backward}^-
&=
  \frac{2\beta}{\left(k+\beta\right)^2}
  \left[
    \partial_z k
  \right]
  {E}_{0,forward}^+
.
\end{align}

Using the small-$\mathscr{Q}$ approximation for $k$,
 we can write 
~
\begin{align}
  \partial_z
  {E}_{0,backward}^-
&=
  \frac{1}{\left(k+\beta\right)^2}
  \left[
    \partial_z
    \mathscr{Q}
  \right]
   e^{-\imath k z}
.
\end{align}
where the exponential part removes any oscillations
 due to the linear part of $\mathscr{Q}$; 
 i.e. if $\mathscr{Q} = \chi {E}$ then 
~
\begin{align}
  \partial_z
  {E}_{0,backward}^-
&=
  \frac{1}{\left(k+\beta\right)^2}
  \left[
    \partial_z\chi
  \right]
.
\end{align}
So here we see that backward evolving fields are only generated
 from forward evolving fields due to changes
 in the underlying conditions 
 (i.e. either material response or pulse properties), 
 but that for the reflection to be strong those changes will have to be
 significant on the order of a wavelength, 
 or be periodic so that phase matching of the the backward wave
 could occur.

\end{document}